\documentclass[a4paper,11pt]{article}
\usepackage{jcappub}
\usepackage[utf8]{inputenc}
\usepackage{amsmath,mathtools}

\usepackage{tensor}
\usepackage{amssymb}
\usepackage{graphicx}
\usepackage{physics}
\usepackage{lipsum}  
\usepackage{bigints}
\usepackage{verbatim}
\usepackage{bbold}
\usepackage[dvipsnames]{xcolor}
\newcommand{\leri}[1]{\left(#1\right)}

\usepackage{subcaption}

\title{Torsional birefringence in metric-affine Chern-Simons gravity: gravitational waves in late-time cosmology}

\author[a,b,1]{S. Boudet,\note{Corresponding author.}}
\author[c]{F. Bombacigno,}
\author[c]{F. Moretti,}
\author[c,d]{Gonzalo J. Olmo,}

\affiliation[a]{Dipartimento di Fisica, Universit\`{a} di Trento,\\Via Sommarive 14, I-38123 Povo (TN), Italy}
\affiliation[b]{Trento Institute for Fundamental Physics and Applications (TIFPA)-INFN,\\Via Sommarive 14, I-38123 Povo (TN), Italy}
\affiliation[c]{Departament de F\'{i}sica Teòrica and IFIC, Centro Mixto Universitat de València - CSIC, Universitat de València, Burjassot 46100, València, Spain}
\affiliation[d]{Universidade Federal do Cear\'a (UFC), Departamento de F\'isica,\\ Campus do Pici, Fortaleza - CE, C.P. 6030, 60455-760 - Brazil.}

\emailAdd{simon.boudet@unitn.it}
\emailAdd{flavio2.bombacigno@uv.es}
\emailAdd{fabio.moretti@ext.uv.es}
\emailAdd{gonzalo.olmo@uv.es}

\abstract{In the context of the metric-affine Chern-Simons gravity endowed with projective invariance, we derive analytical solutions for torsion and nonmetricity in the homogeneous and isotropic cosmological case, described by a flat Friedmann-Robertson-Walker metric. We discuss in some details the general properties of the cosmological solutions in the presence of a perfect fluid, such as the dynamical stability and the emergence of big bounce points, and we examine the structure of some specific solutions reproducing de Sitter and power law behaviours for the scale factor. Then, we focus on first-order perturbations in the de Sitter scenario, and we study the propagation of gravitational waves in the adiabatic limit, looking at tensor and scalar polarizations. In particular, we find that metric tensor modes couple to torsion tensor components, leading to the appearance, as in the metric version of Chern-Simons gravity, of birefringence, characterized by different dispersion relations for the left and right circularized polarization states. As a result, the purely tensor part of torsion propagates like a wave, while nonmetricity decouples and behaves like a harmonic oscillator. Finally, we discuss scalar modes, outlining as they decay exponentially in time and do not propagate.}

\makeatletter
\gdef\@fpheader{}
\makeatother
\begin{document}

\maketitle
\flushbottom

\section{Introduction}
The existence of physical phenomena still evading a convincing explanation within General Relativity (GR) has driven the literature towards the study of extended theories of gravity \cite{DeFelice:2010aj,Nojiri:2010wj,Olmo:2011uz,Cai:2015emx,NOJIRI20171,Krssak:2018ywd,Olmo:2019flu,Cabral:2020fax,Harko:2020ibn,Capozziello:2022lic,Fernandes:2022zrq}, aimed to address open cosmological and astrophysical problems. Considering the wide variety of theories proposed, it is of paramount importance to constrain available models with observations. In the past years, gravitational wave astronomy opened a new paradigm in this direction and recent and upcoming improvements in their detection are expected to offer new observational inputs.

Amidst extended theories of gravity, those characterized by parity violation are receiving growing attention. Symmetries play a crucial role in our current description of nature and among them, parity is well known to be violated in the Standard Model by the weak interaction. Although General Relativity preserves parity, the question whether the gravitational interaction violates it, is still object of debate and efforts have been devoted to investigate the effects of gravitational parity violation. In this regard, over the years some gravitational models have been proposed, dealing with different kinds of modifications to the standard GR action, like Holst and Nieh-Yan extensions \cite{Iosifidis:2020dck,Li:2022vtn}, degenerate higher-order scalar-tensor theories (DHOST) \cite{PhysRevD.97.044034,Qiao:2019wsh}, Chern-Simons modified gravity (CSMG) \cite{Alexander:2009tp,Sulantay:2022sag}, Hořava-Lifshitz gravity \cite{Zhu:2013fja,Gong:2021jgg} and bumblebee models \cite{Kostelecky:2003fs,Bluhm:2004ep,Bailey:2006fd,Bluhm:2007bd,Delhom:2019wcm,Delhom:2020gfv,Delhom:2022xfo}.

{The CSMG, originally formulated by Jackiw and Pi in \cite{Jackiw:2003pm}, is based on the Chern-Simons modification of electrodynamics \cite{Carroll:1989vb}. In this theory, the U(1)-gauge topological Pontryagin density, $\,^{*}FF\equiv \,^{*}F^{\mu\nu}F_{\mu\nu}$, is coupled to a pseudo-scalar field $\theta(x)$ and it is added to the Maxwell Lagrangian, without violating the gauge invariance. Such a modification, however, is responsible for the Lorentz/CPT symmetry violation \cite{Carroll:1989vb}, as it can be appreciated by rewriting the modified term into the Carroll-Field-Jackiw form, i.e. $v_{\mu}\,^{*}F^{\mu\nu}A_{\nu}$, where $v_{\mu}\equiv \partial_{\mu}\theta$ is the axial vector leading to the Lorentz/CPT symmetry breaking. In a wider sense, moreover, it is possible to identify $v_{\mu}$ with one of the coefficients pertaining the Lorentz/CPT violation in the Standard Model Extension (SME) \cite{Colladay:1996iz, Colladay:1998fq, Kostelecky:2003fs}. Then, in a similar vein, CSMG is obtained by adding to the Lagrangian of GR the gravitational Pontryagin density, coupled to $\theta(x)$ and defined by $\,^{*}RR\equiv\,^{*}R^{\mu\nu\alpha\beta}R_{\nu\mu\alpha\beta}$.}

Different versions of CSMG have been explored. A first broad distinction, characterizing any geometric theory of gravity, corresponds to the two possible assumptions regarding the metric and affine structures of the theory. Most of the literature is focused on the purely metric version of CSMG, obtained imposing a symmetric and metric compatible connection. On the other hand, the alternative metric-affine version of the theory, where the metric and the connection are a priori independent, has received little attention, despite being closer to a gauge theory \cite{Hehl:1994ue,Zanelli:2005sa}, since the CS term is built from the connection of the corresponding gauge field. Moreover, most of the work on metric-affine CSMG focuses on theoretical aspects \cite{Deser:2006ht,doi:10.1063/1.529191,PhysRevD.64.084012,CACCIATORI20062523,PhysRevD.78.025002,Cisterna2019}, while observable effects have only been derived in \cite{Boudet:2022wmb}.

A further classification consists in the possibility of including in the action a standard kinetic term for the pseudo-scalar field or not. The terminology commonly used refers to dynamical and non-dynamical CSMG, respectively, and it was established in the context of metric CSMG, where in the non-dynamical case the scalar field is not a proper dynamical degree of freedom and has to be externally prescribed, ultimately yielding issues such as over-constrained systems of equations in black hole settings \cite{PhysRevD.77.064007}, that are only solved in the dynamical theory, where $\theta(x)$ is endowed with a proper dynamics. On the contrary, as it was recently shown in \cite{Boudet:2022wmb}, in metric-affine CSMG the kinetic term for the pseudo-scalar field plays no fundamental role and both the options are actually viable and lead to a consistent theory.

In this work we will consider the metric-affine CSMG model proposed in \cite{Boudet:2022wmb}, where the invariance under projective transformations of the connection is reinstated modifying the Pontryagin density with additional terms depending on the nonmetricity. This is done in the spirit of preventing the presence of dynamical instabilities as a result of breaking the projective symmetry in metric-affine theories of gravity \cite{BeltranJimenez2019,PhysRevD.103.124031}. We remark, however, that in a broader sense dynamical instabilities can always arise due to the appearance of derivatives of higher order for the metric field, just as much as in the metric formulation of CSMG, where the Pontryagin density is ultimately responsible for the appearance in the metric field equations of derivatives of the Riemann tensor, leading to third order derivatives for the metric field. In our case, we managed to demonstrate that such terms can only be introduced by specific components of the purely tensor part of the connection, potentially offering a way for evading this issue by looking at peculiar geometric structures of the theory. In fact, the highly non trivial structure of the connection field equations is one of the main difficulties in finding exact solutions. It can be tackled via a perturbative approach, as in \cite{Boudet:2022wmb}, or exploiting, indeed, the symmetries of the spacetime under consideration, as it is done in this work. In particular, we focus on isotropic and homogeneous cosmological settings, deriving analytical solutions for the connection, which turns out to be expressed in terms of the metric and scalar field. Their behavior is then computed solving the remaining equations, resulting in several, physically viable scenarios.

Among them we examine in more depth the case of a de Sitter expansion phase, studying the propagation of gravitational waves on such background, with the aim of highlighting potentially observable effects. In CSMG, these are usually affected by parity violation, which is expected to leave distinctive features as already well established in contexts such as CMB polarization \cite{ALEXANDER2008444,PhysRevLett.83.1506,Bartolo:2018elp,Bartolo:2017szm}, the baryon asymmetry problem \cite{PhysRevLett.96.081301,PhysRevD.69.023504,Alexander_2006} and black hole perturbations \cite{PhysRevD.80.064008,PhysRevD.81.089903,PhysRevD.81.124021,PhysRevD.80.064006}. The effects of metric CSMG on primordial gravitational waves and inflation have been addressed in \cite{PhysRevD.99.064049,Odintsov:2021kup,sym14040729,PhysRevD.105.104054}.

Parity violation is known to give rise to birefringence phenomena \cite{Conroy:2019ibo,Qiao:2019wsh,Zhao:2019xmm,PhysRevD.98.124018,Chatzistavrakidis:2021oyp,Hohmann:2022wrk,Wu:2021ndf}. In particular, we speak of velocity birefringence when waves with left and right handed polarizations \cite{Isi:2022mbx} propagate with different speeds, while in the presence of a different friction term for the two modes we have amplitude birefringence, in which the enhancement or suppression of the wave depends on its chiral state. Furthermore, when propagation within a material medium is also considered, it can be demonstrated that velocity birefringence can generate amplitude birefringence through the Landau damping phenomenon \cite{Bombacigno:2022naf}, consisting in the kinematic damping of gravitational waves in the absence of collisions (see \cite{Chesters:1973wan,PhysRevD.13.2724,Gayer:1979ff,Weinberg:2003ur,Lattanzi:2005xb,Benini:2010zz,Flauger:2017ged,Baym:2017xvh,Moretti:2020kpp,Moretti:2021ljj,Garg:2021baw,Garg:2022wdm} for more details). Gravitational wave birefringence is a well-established result of metric CSMG \cite{Jackiw:2003pm,Martin-Ruiz:2017cjt,Nojiri:2019nar,Nojiri:2020pqr,Li:2022grj}, and in this work we show for the first time how such a phenomenon is present also in the metric-affine formulation. In particular, we demonstrate how in the adiabatic limit birefringence is induced by the coupling of the gravitational perturbations with the tensor torsion stresses, which as a result dynamically propagate as a wave. Scalar modes, instead, do not exhibit wave propagation in agreement with the results of metric CSMG, and corresponding torsion and nonmetricity components just decay exponentially in time.

The paper is organized as follows. In section \ref{sec2} we briefly review the results of \cite{Boudet:2022wmb} and we discuss in some detail the role of the affine connection in generating dynamical instabilities. In section \ref{sec3} we solve the equations for the affine connection for an isotropic and homogeneous cosmological background, deriving the expressions for its components in terms of the Hubble function and the pseudo-scalar field. In section \ref{sec4} we derive the equations for the metric and the pseudo-scalar field in the presence of a stress energy tensor for a perfect fluid, and we discuss some general properties of the associated Friedmann equation. In section \ref{sec5} we explicitly calculate some analytic solutions, reproducing the well known profiles of the de Sitter and power-law behaviours. Then in section \ref{sec6} we study the propagation of gravitational waves on the de Sitter background, analyzing in some detail the origin and the properties of the birefringence. Finally, in section \ref{sec7} conclusions are drawn and future perspectives outlined.

Spacetime signature is chosen mostly plus and the gravitational coupling set as $\kappa^2=8\pi$, using geometrized units: $G=c=1$. The Levi-Civita tensor $\varepsilon_{\mu\nu\rho\sigma}$  is defined in terms of the completely antisymmetric symbol $\epsilon_{\mu\nu\rho\sigma}$, with $\epsilon_{0123}=1$. For details on the metric-affine formalism and the conventions used we refer the reader to App.~\ref{appendix A}.

\section{Connection in projective invariant CSMG}\label{sec2}
\noindent In this section we briefly review the results of \cite{Boudet:2022wmb}, to which we refer the reader for all the details concerning the metric-affine formulation of Chern-Simons gravity.
The starting point of our analysis is the action
\begin{equation}
S=\frac{1}{2\kappa^2}\int d^4x\sqrt{-g}\leri{\mathcal{R}+\frac{\alpha}{8}\theta(x)\varepsilon\indices{^{\mu\nu\rho\sigma}}\leri{\mathcal{R}\indices{^\alpha_{\beta\mu\nu}}\mathcal{R}\indices{^\beta_{\alpha\rho\sigma}}-\frac{1}{4}\hat{\mathcal{R}}\indices{_{\mu\nu}}\hat{\mathcal{R}}\indices{_{\rho\sigma}}} -\frac{\beta}{2}\nabla_\mu \theta \nabla^\mu\theta},
\label{action CS}
\end{equation}
where $\mathcal{R}\indices{^\mu_{\nu\rho\sigma}}$ is the Riemann tensor of the torsionfull and not metric compatible independent connection, $\hat{\mathcal{R}}_{\mu\nu}=\mathcal{R}\indices{^\mu_{\mu\rho\sigma}}$ is the homotetic curvature and $\theta(x)$ is a pseudo-scalar field (for definitions and notations regarding the metric-affine formalism we refer the reader to Appendix~\ref{appendix A}). Like in the metric formulation, this action contains two parameters $\alpha$ and $\beta$, which control the modified Pontryagin density and the kinetic term for the pseudo-scalar field, respectively. By inspection of \eqref{action CS}, we see that the dimensions of these objects must be the following powers of a length: $[\alpha]=L^A$, $[\theta]=L^{2-A}$, $[\beta]=L^{2A-4}$, with $A$ an arbitrary constant. Note that $\alpha$ can always be reabsorbed in the definition of $\theta$ by the rescaling $\theta\rightarrow\theta/\alpha$ and $\beta\rightarrow\alpha^2 \beta$. This amounts to set $A=0$ and work with $[\theta]=L^2$ and $[\beta]=L^{-4}$, as we will do in section \ref{sec3}.

As discussed in \cite{Boudet:2022wmb}, the action \eqref{action CS} represents the only metric-affine Chern-Simons model endowed with a modified Pontryagin term which remains invariant under the projective transformation
\begin{equation}
\Gamma\indices{^\rho_{\mu\nu}}\rightarrow\tilde{\Gamma}\indices{^\rho_{\mu\nu}}=\Gamma\indices{^\rho_{\mu\nu}}+\delta\indices{^\rho_\mu}\xi_\nu,
    \label{projective}
\end{equation}
and does not spoil topologicity for a constant $\theta(x)$, as it can be verified from:
\begin{equation}
\begin{split}
    &\sqrt{-g}\varepsilon^{\mu\nu\rho\sigma}\leri{\mathcal{R}\indices{^\alpha_{\beta\mu\nu}}\mathcal{R}\indices{^\beta_{\alpha\rho\sigma}}-\frac{1}{4}\hat{\mathcal{R}}_{\mu\nu}\hat{\mathcal{R}}_{\rho\sigma}}=\\
    &=\epsilon^{\mu\nu\rho\sigma}\partial_\mu\leri{\Gamma\indices{^\alpha_{\beta\nu}}\partial_\rho\Gamma\indices{^\beta_{\alpha\sigma}}+\frac{2}{3}\Gamma\indices{^\alpha_{\beta\nu}}\Gamma\indices{^\beta_{\gamma\rho}}\Gamma\indices{^\gamma_{\alpha\sigma}}-\frac{1}{4}\Gamma\indices{^\alpha_{\alpha\nu}}\partial_\rho\Gamma\indices{^\beta_{\beta\sigma}}}.
\end{split}
\end{equation}
Now, by varying \eqref{action CS} with respect to the affine connection we can derive the general equation
\begin{align}
    &-\nabla_\lambda\leri{\sqrt{-g} g^{\mu\nu}}+\delta^\nu_\lambda\nabla_\rho\leri{\sqrt{-g} g^{\mu\rho}}+\sqrt{-g}\leri{g^{\mu\nu}T\indices{^\tau_{\lambda\tau}}-\delta\indices{^\nu_\lambda}T\indices{^{\tau\mu}_\tau}+T\indices{^{\nu\mu}_\lambda}}=\nonumber\\
    &=\frac{\alpha}{2}\sqrt{-g}\,\varepsilon^{\alpha\beta\gamma\nu}\leri{\mathcal{R}\indices{^\mu_{\lambda\beta\gamma}}-\frac{1}{4}\delta\indices{^\mu_\lambda}\hat{\mathcal{R}}\indices{_{\beta\gamma}}}\nabla_\alpha\theta,
    \label{equation connection general}
\end{align}
where $\nabla$ stands for the general covariant derivative defined from the entire connection. From \eqref{equation connection general}, by taking into account the decomposition of torsion and nonmetricity in their irreducible components and setting the trace of torsion to zero by virtue of the projective invariance (see the discussion in \cite{Boudet:2022wmb} and Appendix \ref{appendix A}), it is possible to extract the equations for the vector and tensor part, which can be arranged respectively as 
\begin{align}
    4P_\mu-Q_\mu&=\frac{\alpha}{2}\leri{\varepsilon^{\alpha\beta\gamma\delta}\leri{\mathcal{R}\indices{_{\mu\beta\gamma\delta}}+\mathcal{R}\indices{_{\beta\mu\gamma\delta}}}+\frac{1}{2}\varepsilon\indices{_\mu^{\alpha\beta\gamma}}\hat{\mathcal{R}}\indices{_{\beta\gamma}}}\nabla_\alpha\theta,
    \label{symmetric part}\\
    P_\mu-Q_\mu&=\frac{\alpha}{2}\varepsilon^{\alpha\beta\gamma\delta}\leri{\mathcal{R}\indices{_{\mu\beta\gamma\delta}}-\mathcal{R}\indices{_{\beta\mu\gamma\delta}}}\nabla_\alpha\theta,
    \label{antisymmetric part}\\
    S_\mu&=\alpha\leri{\mathcal{R}\indices{^\rho_\mu}+\mathcal{R}\indices{^{\rho\sigma}_{\mu\sigma}}-\delta\indices{^\rho_\mu}\mathcal{R}}\nabla_\rho\theta,
    \label{equation component connection s proj}
\end{align}
and
\begin{equation}
\begin{split}
q_{\nu\mu\lambda}-\Omega_{\lambda\mu\nu}=&\frac{\alpha}{2}\,\varepsilon\indices{^{\alpha\beta\gamma}_\nu}\leri{\mathcal{R}\indices{_{\mu\lambda\beta\gamma}}-\frac{1}{4}g\indices{_{\mu\lambda}}\hat{\mathcal{R}}\indices{_{\beta\gamma}}}\nabla_\alpha\theta
-\dfrac{1}{6} \varepsilon_{\nu\mu\lambda\sigma}S^{\sigma}+\\
&-\frac{1}{9}\leri{g_{\mu\nu}\leri{2Q_\lambda+P_{\lambda}}-g_{\nu\lambda}\leri{4Q_\mu-7P_\mu}+g_{\mu\lambda}\leri{\frac{1}{2}Q_\nu-2P_\nu}}.
\label{equation connection tensor part general solved}
\end{split}
\end{equation}
In the following, it will be useful to separate \eqref{equation connection tensor part general solved} into its symmetric and antisymmetric parts in the indices $\mu,\,\lambda$, in order to deal as in \eqref{symmetric part}-\eqref{antisymmetric part} with the symmetric and antisymmetric part of the Riemann tensor in its first two indices, i.e.
\begin{equation}
\begin{split}
\Omega_{(\lambda\mu)\nu}=&-\frac{\alpha}{2}\,\varepsilon\indices{^{\alpha\beta\gamma}_\nu}\leri{\mathcal{R}\indices{_{(\mu\lambda)\beta\gamma}}-\frac{1}{4}g\indices{_{\mu\lambda}}\hat{\mathcal{R}}\indices{_{\beta\gamma}}}\nabla_\alpha\theta+\\
&+\frac{1}{9}\leri{g_{\mu\nu}\leri{4P_{\lambda}-Q_\lambda}+g_{\lambda\nu}\leri{4P_{\mu}-Q_\mu}-\frac{1}{2}g_{\mu\lambda}\leri{4P_{\nu}-Q_\nu}}
\label{equation connection tensor part symm}
\end{split}
\end{equation}
and
\begin{equation}
\begin{split}
q_{\nu\mu\lambda}-\Omega_{[\lambda\mu]\nu}=&\frac{\alpha}{2}\,\varepsilon\indices{^{\alpha\beta\gamma}_\nu}\leri{\mathcal{R}\indices{_{[\mu\lambda]\beta\gamma}}}\nabla_\alpha\theta+\\
&-\frac{1}{6}\varepsilon_{\nu\mu\lambda\sigma}S^\sigma-\frac{1}{3}\leri{g_{\mu\nu}\leri{Q_\lambda-P_{\lambda}}-g_{\lambda\nu}\leri{Q_\mu-P_{\mu}}}.
\label{equation connection tensor part antisymm}
\end{split}
\end{equation}
The advantage of such a decomposition relies on the possibility of isolating the purely metric part of the Riemann tensor, which according to \eqref{perturbative expansion riemann} results encoded in the antisymmetric component of the first two indices, i.e.
\begin{equation}
    \mathcal{R}_{[\mu\rho]\nu\sigma}=R_{\mu\rho\nu\sigma}+A\indices{_{[\mu\rho]\nu\sigma}},
\end{equation}
where we denoted with $A_{\mu\rho\nu\sigma}$ the non-Riemannian terms depending on the distorsion tensor. The symmetric part, instead, can be rewritten as
\begin{equation}
    \mathcal{R}_{(\mu\rho)\nu\sigma}=\nabla_{[\nu}Q_{\sigma]\mu\rho}-\frac{1}{2}T\indices{^\tau_{\nu\sigma}}Q_{\tau\mu\rho},
\end{equation}
which can be obtained from the commutator of the covariant derivatives
\begin{equation}
    [\nabla_\nu,\nabla_\sigma]t_{\mu\rho}=-\mathcal{R}\indices{^\tau_{\mu\nu\sigma}}t_{\tau\rho}-\mathcal{R}\indices{^\tau_{\rho\nu\sigma}}t_{\mu\tau}+T\indices{^\tau_{\nu\sigma}}\nabla_\tau t_{\mu\rho},
\end{equation}
once we identify the generic tensor $t_{\mu\rho}$ with the metric field $g_{\mu\rho}$ and we take into account the definition of nonmetricity.

\subsection{The role of the affine connection in dynamical instabilities}
Dynamical instabilities can arise if derivatives of order higher than two appear in the equations of motion \cite{Bluhm:2004ep,Woodard:2006nt,Gleyzes:2014dya,Langlois:2015cwa,Aoki:2019rvi,Baldazzi:2021kaf,Percacci:2020ddy}, and it is known, for instance, that metric CSMG is fully coherent only in the limit of small coupling $|\alpha^2/\beta| \ll 1$, where the Pontryagin density is subdominant with respect to the kinetic term for $\theta(x)$ \cite{PhysRevD.91.024027}. This is due to the fact that the Chern-Simons modification is responsible for an additional contribution in the equation of the metric field, the so-called C-tensor \cite{Jackiw:2003pm,Alexander:2009tp}, which containing derivatives of the Ricci tensor, causes the presence of third order derivatives of $g_{\mu\nu}(x)$ in the field equations. In our case, instead, by varying \eqref{action CS} with respect to the metric, we simply obtain
\begin{equation}
    \mathcal{R}_{(\mu\nu)}-\frac{1}{2}g_{\mu\nu}\mathcal{R}=\kappa^2 T_{\mu\nu}+\frac{\beta}{2} \left(\nabla_\mu \theta \nabla_\nu \theta- \frac{1}{2} g_{\mu\nu} \nabla_\rho \theta \nabla^\rho \theta \right),
\end{equation}
which by taking into account \eqref{perturbative expansion riemann} can be still rearranged in a form resembling its purely metric counterpart, i.e.
\begin{equation}
    G_{\mu\nu}+C_{\mu\nu}=\kappa^2 T_{\mu\nu}+\frac{\beta}{2} \left(\nabla_\mu \theta \nabla_\nu \theta- \frac{1}{2} g_{\mu\nu} \nabla_\rho \theta \nabla^\rho \theta \right)\equiv \kappa^2 T^{TOT}_{\mu\nu},
\end{equation}
where the C-tensor is now identified with the metric-affine terms containing the distorsion tensor (see \eqref{c tensor affine}).
\\A careful analysis of \eqref{c tensor affine} allows us to identify the origin of possible dynamical instabilities in the dependence of the distorsion tensor on second order derivatives of the metric and pseudo-scalar fields. 
Then, if we look back at the equations for the connection, we immediately see that $\theta(x)$ only appears by means of its first derivative, so that dynamical instabilities can be only caused by the metric field, that is to say by metric Riemann tensor terms potentially appearing in \eqref{equation connection general}. In particular, it is easy to observe that such terms can be always neglected in the equations for the vectors components. In \eqref{symmetric part}-\eqref{antisymmetric part}, indeed, the metric Riemann tensor term cancels out by virtue of the Bianchi identity, while the equation for the axial trace, by using the equation for the metric field, can be recast as
\begin{equation}    S_\mu=\alpha\leri{2\kappa T^{TOT}_{\rho\mu}+\leri{\cdots}}\partial^\rho\theta,
\label{solution s matter}
\end{equation}
where the dots stand for contributions proportional to the distorsion tensor $N\indices{^\rho_{\mu\nu}}$ and its metric covariant derivative. It follows, therefore, that in vacuum, where no additional interaction terms are introduced at the effective level by the axial trace, the metric Riemann tensor only survives in \eqref{equation connection tensor part antisymm}, which is ultimately responsible for the presence or not of dynamical instabilities in metric-affine CSMG. Actually, we note that a non vanishing stress energy tensor does not necessarily imply the emergence of instabilities. Indeed, if the tensor $T_{\mu\nu}$ contains only first covariant derivatives of the field, as it occurs for example for the standard scalar or electromagnetic cases, the purely affine components of the Riemann tensor can generate in the equation of motion for $g_{\mu\nu}$ and $\theta(x)$ at most second order derivatives. On the other hand, second derivatives of the matter fields featuring the $T_{\mu\nu}$ would already imply third order derivatives in the equations of those fields, representing an issue in itself, unrelated to the setting of metric-affine CSMG.

These considerations can help us in designing strategies for ensuring the absence of higher-order derivatives in the field equations. This is the same requirement needed in the metric CSMG, as mentioned at the beginning of this section. However, the crucial difference with respect to metric CSMG is that in that case higher-order derivatives directly appear in the metric field equation, while in the metric-affine theory they are introduced via the affine contributions only if the latter depend on them. This additional step allows for a new approach to this problem that is not available in metric CSMG, which consists on acting previously on the kinematic structure of the theory. This can be achieved, for instance, by considering from the very beginning simplified metric-affine geometries, where the rank-3 tensor part of torsion and nonmetricity is neglected, and we only deal with vector components. Such an assumption, indeed, automatically gets rid of \eqref{equation connection tensor part antisymm}, leaving us only with the set of equations for the nonmetricity vectors \eqref{symmetric part} - \eqref{antisymmetric part}, and the axial trace of torsion \eqref{equation component connection s proj}. It represents the minimal prescription for preserving projective invariance, given that $q_{\rho\mu\nu}$ and $\Omega_{\rho\mu\nu}$ are unaffected by \eqref{projective}, and the requirement they are vanishing does not spoil in principle the hypothesis of \eqref{action CS}. In this approach, higher-order derivatives would be absent at the full unperturbed and background independent level, without imposing restrictions on the parameters of the theory.

If instead the tensorial parts are considered, one is forced to resort to approaches analogue to the ones adopted in the metric theory. For instance, one can consider metric-affine CSMG only as an effective theory in the small coupling limit $\alpha\ll 1$ ($\alpha \theta \ll 1$, after the rescalings performed in section \ref{sec3}), thus making the first term on the right-hand side of \eqref{equation connection tensor part antisymm} negligible, ensuring the absence of higher-order derivatives at the effective level. The same outcome is also obtained in the linearized setting already adopted in \cite{Boudet:2022wmb}, where the perturbative expansion of the scalar field is performed on a constant background.

Finally, a viable strategy consists in seeking for spacetime configurations where the components of the metric Riemann tensor containing second order derivatives of the metric field are prevented from appearing in \eqref{equation connection tensor part antisymm} by the symmetries of the problem, which restricting the possible dependence of $\theta(x)$ on specific spacetime coordinates, also selects the components of the Riemann tensor. The equations describing the evolution of the connection reduce then to a highly coupled system of first order differential equations for the metric-affine components, which contain at most first derivatives of the pseudo-scalar field and of the metric tensor, by means of the Levi-Civita connection appearing in the metric covariant derivatives. 

Even though the role of matter in generating instabilities evades the purpose of this work and it calls for further investigations, we remark that the previous conclusion relies on the assumption that the connection does not couple to matter. In fact, in the general case of a non vanishing hypermomentum \cite{Hehl:1976kv,Iosifidis:2020gth}
\begin{equation}
    \Delta\indices{_\lambda^{\mu\nu}}\equiv -\frac{2}{\sqrt{-g}}\frac{\delta S_M}{\delta\Gamma\indices{^\lambda_{\mu\nu}}}\neq 0,
\end{equation}
where $S_M$ denotes collectively the Lagrangian for the matter fields, we expect contributions analogous to the one in \eqref{solution s matter}, in terms of the hypermomentum components, also in the equations for the other parts of $\Gamma\indices{^\lambda_{\mu\nu}}$, so that no conclusive arguments can be drawn in this situation. In the following we discuss in detail the case of Friedman-Lemaitre-Robertson-Walker (FLRW) cosmology, where the symmetries of the spacetime enable us to derive analytical solutions devoid of instabilities both for the metric and affine part.

\section{Solving the connection in FLRW spacetimes}\label{sec3}
Let us consider a FLRW line element in spherical coordinates, i.e.
\begin{equation}
    ds^2 = - dt^2 +a^2(t) \left( \frac{dr^2}{1-k r^2} + r^2 (d\psi^2+\sin^2\psi d\varphi^2) \right).
\end{equation}
For such a configuration, second-order derivatives of the metric are displayed only in the Riemann components
\begin{equation}
    R_{trtr}=-\frac{a\ddot{a}}{1-kr^2},\;R_{t\psi t\psi}=-r^2 a\ddot{a},\;R_{t\varphi t\varphi}=-r^2 \sin^2\psi a\ddot{a}.
    \label{riemann frw}
\end{equation}
Now, since $\theta=\theta(t)$, \eqref{riemann frw} can never appear in \eqref{equation connection tensor part antisymm}, where we simply retain the purely spatial part of the Riemann tensor, i.e.
\begin{equation}
    \frac{\dot{\theta}}{2}\,\varepsilon\indices{^{tij}_k}R\indices{_{ijij}},
\end{equation}
which contains at most first derivatives of the metric function $a(t)$. Then, as discussed in \cite{Hohmann:2019fvf,Iosifidis:2020gth}, isotropy and homogeneity allow us to rewrite the connection in terms of a set of five scalar functions $\{K_1(t),...,K_5(t)\}$, which entirely define the affine structure of the spacetime (see also \cite{Bahamonde:2020fnq,Bahamonde:2022kwg} for exact black hole solutions in metric-affine models). In particular, in the flat $k=0$ case, the vector and tensor components of the connection result in
\begin{align}
    T^{t} &= 3(K_3-K_4),\label{First affine components FRW}\\
    S^{t} &= -\frac{12 K_5}{a},\\
    Q^{t} &= 2K_1 + 6K_4 - 6 H,\\
    P^{t} &= 2K_1 - 3 \frac{K_2}{a^2} + 3 K_3,\\
    \Omega_{ttt} &= -K_1 + K_3 + K_4 - \frac{K_2}{a^2} - H,\\
    \Omega_{t\psi\psi} &= \Omega_{\psi\psi t} = r^2 \Omega_{trr},\\
    \Omega_{t\varphi\varphi} &= \Omega_{\varphi\varphi t} = r^2 \text{sin}^2\psi \Omega_{trr},\\
    \Omega_{trr} &= \Omega_{rrt}=\frac{a^2}{3}\Omega_{ttt},\\
     q_{\mu\nu\rho} &= 0,
    \label{affine components FRW}
\end{align}
where $H=\dot{a}/a$ is the usual Hubble function. We immediately see that having imposed the condition $T_{\mu} =0$ by virtue of the projective invariance, leads us to the constraint $K_4(t)= K_3 (t)$. With a bit of work, \textbf{two solutions} to \eqref{equation connection general} can be obtained and the form of the connection is completely determined in terms of the metric functions and scalar field\footnote{From now on we set $\alpha=1$ and therefore $[\theta]=L^2$, $[\beta]=L^{-4}$ (see section \ref{sec2}).}:
\begin{align}
    K_1(t) &= - \frac{H K_5\dot{\theta}}{a+K_5 \dot{\theta}} ,\\
    K_2(t) &=  \frac{a^3 H}{a+K_5 \dot{\theta}},\\
    K_3(t) &= K_4 =  \frac{a H}{a+K_5 \dot{\theta}},\\
    K_5(t) &= \frac{a}{\dot{\theta}} \left( -1+\epsilon\,\sqrt{\frac{1+\sqrt{1+4 H^2 \dot{\theta}^2}}{2}} \right),
\end{align}
where we assumed $a+K_5 \dot{\theta}\neq 0$ and we introduced the parameter $\epsilon=\pm 1$ in order to label the different solutions. We note that for $a+K_5 \dot{\theta}= 0$, the connection equation yields the condition $H(t)\equiv0$, which corresponds to a static solution where the scale factor does not evolve in time. Eventually, inserting the solutions for $\{K_1,...,K_5\}$ in \eqref{First affine components FRW}-\eqref{affine components FRW}, we end up with
\begin{align}
    S^t &= \frac{12 K_5}{a}= \frac{12}{\dot{\theta}} \left( -1+ \epsilon\,\sqrt{\frac{1+\sqrt{1+4 H^2 \dot{\theta}^2}}{2}} \right)
    \label{solution S frw},\\
    Q^t &= 4P^t = -8K_1 = 8H \left( 1-\epsilon\,\sqrt{\frac{2}{1+\sqrt{1+4 H^2 \dot{\theta}^2}}} \right),
    \label{solution PQ frw}
\end{align}
while all other affine components are vanishing. In particular, the tensor part of the connection is identically vanishing and for a FLRW spacetime we only deal with the vector part, which is characterized by a Weyl geometry configuration, where $Q\indices{_{\rho\mu\nu}} = P_\rho g_{\mu\nu}$ (see \eqref{non metricity decomposition}).

\section{Metric and scalar field equations}\label{sec4}
We consider now the equation for the metric field, which as observed in sec.~\ref{sec2} can be rewritten as
\begin{equation}
    G_{\mu\nu} + C_{\mu\nu} =\frac{\beta}{2} \left(\nabla_\mu \theta \nabla_\nu \theta- \frac{1}{2} g_{\mu\nu} \nabla_\rho \theta \nabla^\rho \theta \right)+ \kappa T_{\mu\nu},
    \label{equation connection non pert}
\end{equation}
where $T_{\mu\nu}$ is the stress energy tensor for a perfect fluid:
\begin{equation}
    T\indices{_\mu_\nu} = (\rho + p ) u_\mu u_\nu + p g_{\mu\nu}.
\end{equation}
For a purely vector connection, endowed with a Weyl structure, the C-tensor assumes the form
\begin{equation}\label{Cotton}
    C_{\mu\nu} = -\bar{\nabla}_{(\mu} P_{\nu)} + g_{\mu\nu}\bar{\nabla}_\rho P^\rho + \frac{1}{2}\left( P_\mu P_\nu + \frac{1}{2}g_{\mu\nu} P_\rho P^\rho \right) - \frac{1}{72}\left( S_\mu S_\nu + \frac{1}{2}g_{\mu\nu} S_\rho S^\rho \right),
\end{equation}
and substituting the solution for the connection in terms of the $K_i(t)$, we obtain its non vanishing components:
\begin{align}
  C\indices{^t_t} &=3\leri{\frac{ K_5}{a}}^2+3H^2\leri{1-\leri{\frac{a}{a+K_5 \dot{\theta}}}^2},\\
   C\indices{^r_r} &=\leri{\frac{ K_5}{a}}^2+(2\dot{H}+3H^2)\leri{\frac{K_5\dot{\theta}}{a+K_5\dot{\theta}}}^2+\frac{2\frac{d}{dt}(aH\dot{\theta}K_5) }{a^2 \left(a+K_5 \dot{\theta}\right)^2},\\
   C\indices{^\theta_\theta} &= C\indices{^\varphi_\varphi}=C\indices{^r_r}.
\end{align}
More interestingly, the C-tensor can also be written as the effective stress energy tensor for a perfect fluid with energy density $\rho_{eff}$ and pressure $p_{eff}$, by the identification
\begin{equation}
T^{eff}_{\mu\nu} =-\frac{C_{\mu\nu}}{\kappa}\equiv (\rho_{eff} + p_{eff} ) u_\mu u_\nu + p_{eff} g_{\mu\nu},
\end{equation}
where by comparison with \eqref{solution S frw}-\eqref{solution PQ frw} it is possible to obtain the following expressions for density and pressure, i.e.
\begin{align}
\rho_{eff} &=-\frac{1}{\kappa}\leri{ \frac{3}{4} (P^t)^2 -\frac{1}{48} (S^t)^2 - 3H P^t},\\
p_{eff} &=-\frac{1}{\kappa} \leri{\frac{1}{144} (S^t)^2 - \frac{1}{4}( P^t )^2 +2HP^t+ \dot{P}^t},
\end{align}
whose ratio defines the effective polytropic index
\begin{equation}
w_{eff} = \frac{p_{eff}}{\rho_{eff}}.
\end{equation}
This interpretation of the C-tensor allows us to rewrite \eqref{equation connection non pert}, by setting $u^\mu=(1,0,0,0)$, as
\begin{align}
   3H^2 &= \kappa \tilde{\rho} + \frac{\beta}{4}  \dot{\theta}^2 ,
   \label{friedmann rho totale}\\
   3 H^2 + 2 \dot{H}&= -\kappa \tilde{p} - \frac{\beta}{4}  \dot{\theta}^2,
\end{align}
where
\begin{equation}
    \tilde{\rho}\equiv \rho+\rho_{eff},\qquad \tilde{p}\equiv p+p_{eff}.
\end{equation}
It is therefore clear that torsion and nonmetricity contribute to the total amount of energy concretely determining the evolution of the scale factor, so that it seems reasonable to demand for the positiveness of $\tilde{\rho}$, rather than merely for $\rho$ (see the discussion at the end of this section and in section \ref{sec5}).

Now, regarding the equation for the pseudo-scalar field, we start by varying \eqref{action CS} with respect to $\theta(x)$, which gives us
\begin{align}
    \beta\Box\theta + \frac{1}{8}    \varepsilon\indices{^{\mu\nu\rho\sigma}}\leri{\mathcal{R}\indices{^\alpha_{\beta\mu\nu}}\mathcal{R}\indices{^\beta_{\alpha\rho\sigma}}-\frac{1}{4}\mathcal{R}\indices{^\alpha_{\alpha\mu\nu}}\mathcal{R}\indices{^\beta_{\beta\rho\sigma}}}=0
    \label{equation scalar field nonperturb},
\end{align}
which once we take into account the affine structure results in:
\begin{equation}
    \beta\leri{\ddot{\theta}+3H\dot{\theta}}+  \mathcal{D}_1(a,\dot{a},\dot{\theta},\ddot{\theta})+ \mathcal{D}_2(a,\dot{a},\ddot{a},\dot{\theta})=0,
\end{equation}
where the functions $\mathcal{D}_1,\,\mathcal{D}_2$ are given by
\begin{align}
    \mathcal{D}_1(a,\dot{a},\dot{\theta},\ddot{\theta})&\equiv 6\dot{K}_5\left(  \frac{2 a^2 H^2 }{(a+  K_5\dot{\theta})^3}-\frac{ a H^2 }{(a+  K_5\dot{\theta})^2}-\frac{ K_5^2 }{a^3}\right)  -\frac{12   a H^2 K_5^2}{(a+  K_5\dot{\theta})^3}\ddot{\theta},\\
    \mathcal{D}_2(a,\dot{a},\ddot{a},\dot{\theta})\equiv&  12 a H K_5\left( \frac{    2H^2+\dot{H}   }{(a+  K_5\dot{\theta})^2} -\frac{ a H^2 }{(a+  K_5\dot{\theta})^3} \right) .
\end{align}
As already observed in \cite{Boudet:2022wmb}, in the limit $\beta\to 0$ the equation for $\theta(t)$ is still dynamical, as opposed to what occurs in the metric formulation of the CSMG, where in the absence of the kinetic term for the pseudo-scalar field one deals with an over-constrained metric theory (see \cite{PhysRevD.77.064007}).
Even if the explicit form of the C-tensor and the pseudo-scalar field equation are quite involved, several analytical solutions can be still derived for the case $\beta=0$, which remarkably admits GR known solutions for the scale factor evolution in the presence of non trivial energy density profiles. In particular, by assuming pressure and energy density related by the equation of state $p=w \rho$, where $w$ denotes here the polytropic index for the perfect fluid quantities, and since it is possible to check that the stress energy tensor is conserved throughout the solutions, which is to say that the C-tensor is covariantly conserved, i.e. $\tensor[^{(L)}]{\nabla}{_{\mu}} C^{\mu\nu} = 0$, we can still express the energy density as $\rho=\frac{\rho_0}{a^{3(1+w)}}$, with $\rho_0$ a constant. It is possible to show, moreover, that the relation $a+K_5 \dot{\theta}\neq 0$ is preserved by the dynamics. Then, with a bit of manipulation, we can rearrange the $tt$ component of the metric equation in the canonical form
\begin{equation}
    H^2=\frac{\leri{2\leri{1+\frac{\dot{\theta}^2}{6}\leri{\kappa\rho+\frac{\beta}{4}\dot{\theta}^2}}^2-1}^2-1}{4\dot{\theta}^2},
\end{equation}
which represents the Friedmann equation for the metric-affine CSMG in the presence of a perfect fluid, and where with respect to \eqref{friedmann rho totale} we made explicit the dependence of $\tilde{\rho}$ on the scale factor and the pseudo-scalar field. Then, by requiring $H^2\ge0$, we end up with the conditions
\begin{equation}
    \kappa\rho\le-\frac{48+\beta\dot{\theta}^4}{4\dot{\theta}^2}\quad\cup\quad \kappa\rho=-\frac{24+\beta\dot{\theta}^4}{4\dot{\theta}^2}\quad\cup\quad \kappa\rho\ge-\frac{\beta\dot{\theta}^2}{4},
    \label{big bounce existence}
\end{equation}
and we see that for $\beta\ge0$, where no ghost instabilities can arise in \eqref{action CS}, there exist domains for the energy density of the perfect fluid where it apparently takes negative values, i.e. $\rho_0<0$. However, as formerly outlined and furtherly discussed in section \ref{sec5}, the contribution of torsion and nonmetricity can restore the correct sign for the total energy density. In addition, making explicit the dependence of $\rho$ on the scale factor, it is possible to derive from \eqref{big bounce existence} the values of the scale factor where the Hubble function vanishes, i.e.
\begin{equation}
    a_B=\left\{\leri{-\frac{48+\beta\dot{\theta}^4}{4\kappa\rho_0\dot{\theta}^2}}^{-\frac{1}{3(1+w)}},\leri{-\frac{24+\beta\dot{\theta}^4}{4\kappa\rho_0\dot{\theta}^2}}^{-\frac{1}{3(1+w)}},\leri{-\frac{\beta\dot{\theta}^2}{4\kappa\rho_0}}^{-\frac{1}{3(1+w)}}\right\}_{\dot{\theta}=\dot{\theta}_B}.
    \label{a bounce}
\end{equation}
Even if it goes beyond the scope of this work, here we just note that they represent possible big bounce points, characterizing the evolution of the scale factor in the context of the CSMG, provided configurations with $\rho_0<0$, besides those ones discussed in section \ref{sec5}, be feasible.

\section{Analytic solutions}\label{sec5}
In this section we discuss in more detail the analytic solutions which can be obtained neglecting the kinetic term for the pseudo-scalar field $\theta$, so that all the solutions are to be intended for\footnote{We note that the last of \eqref{a bounce} predicts the possibility of a big bounce still singular, i.e. $a_B=0$.} $\beta=0$. Moreover, for each solution we have to select one of the two solutions for the connection labeled by $\epsilon$. In particular, the first two cases presented below are solutions to the equations only for $\epsilon=-1$, while $\epsilon=+1$ leads to inconsistencies. In the third case the situation is analogous but with the consistent solution now identified by $\epsilon=+1$.
\subsection{De-Sitter phase of acceleration}\label{solution 1}
Here we set $w=-1$, corresponding to $\rho=\rho_0$, and we select $\epsilon=-1$. Under these assumptions it is possible to derive the following solutions for the scale factor and the pseudo-scalar field:
\begin{align}
    a(t) & = a_0 e^{t \sqrt{\Lambda/3}},
    \label{scale factor de sitter CSMG}\\
    \theta(t) &= \pm \frac{6}{\sqrt{\Lambda}} t + \theta_0,
\end{align}
where $a_0$, $\theta_0$ are arbitrary constants and the value of $\rho_0$ is determined by
\begin{equation}
    \rho_0 = -\Lambda /2\kappa.
\label{bare energy density}
\end{equation}
We note that since $\Lambda>0$, \eqref{bare energy density} corresponds actually to a negative bare energy density $\rho_0$. In GR instead, where the C-tensor is vanishing, solution \eqref{scale factor de sitter CSMG} would be obtained for $\rho_0=\frac{\Lambda}{\kappa}>0$. Such a discrepancy is due to the fact that in CSMG torsion and nonmetricity are responsible for an additional contribution to the total energy density and, in particular, they depend on the cosmological constant in the following way:
\begin{align}
    S^t &= \mp 6\sqrt{\Lambda},\\
    Q^t &= 4P^t = 4\sqrt{3\Lambda}.
\end{align}
In this case, indeed, by evaluating explicitly the energy density and pressure associated to the C-tensor, we obtain, respectively
\begin{equation}
    \rho_{eff}=\frac{3\Lambda}{2\kappa},\qquad p_{eff}=-\frac{3\Lambda}{2\kappa},
\end{equation}
which are consistent with a perfect fluid described by a cosmological constant, i.e. $w_{eff}=-1$.

\subsection{Power law solutions reproducing radiation and matter dominated eras}\label{solution 2}
Here we discuss power law solutions for $\epsilon=-1$, which are displayed by
\begin{align}
    a(t) &= a_0 t^m,\\
    \theta(t) &= \pm \frac{\sqrt{3}}{m} t^2 + \theta_0,\\
    \rho(t) &= -\frac{3m^2}{2\kappa} \frac{1}{t^2},
\end{align}
where $a_0$, $\theta_0$ are arbitrary constants and the parameter $m$ is related to the polytropic index by
\begin{equation}
w = \frac{2}{3m}-1.
\label{relation m w}
\end{equation}
Torsion and nonmetricity vectors are then given by
\begin{align}
    S^t &= \mp  \frac{6 \sqrt{3} m}{t},\\
    Q^t &= 4 P^t = \frac{12m}{t},
\end{align}
and they decay linearly in time.
We note that by choosing properly the value of $m$ we can reproduce for the scale factor the well-known behaviours of the radiation and matter dominated eras, i.e. $m_{rad}=1/2$ and $m_{mat}=2/3$, corresponding respectively to the GR values  $w=1/3$ and $w=0$, which are correctly reproduced by \eqref{relation m w}. Also in this case, moreover, the bare energy density turns out to be negative for all values of $m$ and always decaying as the square of the time in agreement with the power law behaviour. Then, in analogy with the discussion of subsec.~\ref{solution 1}, it can be shown that the C-tensor is responsible for the missing energy density, and the evaluation of $\rho_{eff}$ and $p_{eff}$ gives us
\begin{equation}
    \rho_{eff}=\frac{9m^2}{2\kappa t^2},\qquad p_{eff}=-\frac{m}{\kappa t^2}\leri{3-\frac{9m}{2}},
\end{equation}
resulting in the effective polytropic index $w_{eff}=\frac{2}{3m}-1$, in agreement with \eqref{relation m w}.
\subsection{Solution reproducing linear growth of the scale factor}\label{solution 3}
A power law solution with $m=1$ exists also for $\epsilon=1$, in the peculiar case $p=-\rho/3$. It is displayed by:
\begin{align}
    a(t) &= a_0 t,\\
    \theta(t) &= \theta_1 t^2 + \theta_0,\\
    \rho(t) &= \frac{3}{2\kappa \theta_1^2}\left(  -1 + \frac{1}{\sqrt{2}}\sqrt{1+\sqrt{1+16 \theta_1^2}}\right) \frac{1}{t^2},
\end{align}
where $a_0$, $\theta_0$ and $\theta_1$ are arbitrary constants. In this case, moreover, it results that $\rho>0$. Therefore, we have
\begin{align}
    S^t &=  \frac{6}{\theta_1} \left( -1 +\frac{1}{\sqrt{2}}\sqrt{1+\sqrt{1+16\theta_1^2}} \right) \frac{1}{t},\\
    Q^t &= 4P^t = 8 \left( 1-\frac{\sqrt{2}}{\sqrt{1+\sqrt{1+16\theta_1^2}}} \right)\frac{1}{t}.
\end{align}
We conclude this section by noting that for $\beta=0$ relations \eqref{big bounce existence} boil down to
\begin{equation}
    \kappa\rho\le-\frac{12}{\dot{\theta}^2}\quad\cup\quad \kappa\rho=-\frac{6}{\dot{\theta}^2}\quad\cup \quad\kappa\rho\ge 0,
\end{equation}
and it is quite immediate to check that all the solutions previously described comply with these requirements.
\section{Perturbations on FLRW background}\label{sec6}
In this section we are interested in the propagation of gravitational waves on a de Sitter background, described by the exact solution discussed in Sec. \ref{solution 1}. Following the convention of \cite{Nojiri:2019nar,Nojiri:2020pqr}, we adopt for the metric and the pseudo-scalar field the following perturbative expansion:
\begin{equation}
    g_{\mu\nu} = \bar{g}_{\mu\nu} + h_{\mu\nu}, \qquad \theta = \bar{\theta} + \delta \theta,
    \label{gw perturbations}
\end{equation}
where bar quantities denote the results of section \ref{solution 1}. In particular, by choosing the gauge condition $h_{t\mu}=0$ and assuming a wave propagating along the $z$-axis, the purely tensor modes are described by
\begin{equation}
    \delta\theta = 0,\qquad h_{ij} = \begin{pmatrix}
 h_+ & h_\times & 0 \\
 h_\times & -h_+ & 0 \\
 0 & 0 & 0
\end{pmatrix},
\end{equation}
where $h_+$ and $h_\times$ are functions of $t$ and $z$ alone. Regarding the affine sector, we expand the connection as
\begin{equation}
    \Gamma\indices{^\rho_{\mu\nu}}=\bar{\Gamma}\indices{^\rho_{\mu\nu}}+\delta\Gamma\indices{^\rho_{\mu\nu}},
\end{equation}
where in $\delta\Gamma\indices{^\rho_{\mu\nu}}$ are included, in principle, the perturbations of all the irreducible components, namely the tensors $\delta q_{\mu\nu\rho}$ and $\delta \Omega_{\mu\nu\rho}$, which we assume to preserve the symmetries of the unperturbed rank-3 tensors, and the vector parts $\delta S^{\mu}$, $\delta P^{\mu}$ and $\delta Q^{\mu}$. As in \eqref{gw perturbations}, all the connection components are considered only functions of $t$ and $z$. Then, by considering the first order of the linearized equations of motion for the connection, it can be verified that several components of the rank-3 tensor perturbations can be algebraically related, leading to:
\begin{align}
    \delta q_{110} &= \delta q_{202}, \qquad  \delta q_{102} = \delta q_{201}, \qquad  \delta q_{232} = \delta q_{113}, \qquad  \delta q_{123} = \delta q_{213},\\
    \delta\Omega_{111} &= -\delta\Omega_{122} =  -\delta\Omega_{212} = \frac{3}{2a\sqrt{\Lambda}} \delta\Omega_{112}', \qquad \delta\Omega_{211} =  -\delta\Omega_{222} =  \delta\Omega_{112},
\end{align}
where temporal and spacial indices are denoted by $0$ and $1,2,3$, respectively while primes represent derivatives with respect to $z$. In particular, the tensor part of nonmetricity is then determined in its spatial dependence by the harmonic oscillator equation
\begin{equation}
    \delta\Omega_{112}'' + \frac{4\Lambda}{9} a^2 \delta\Omega_{112} = 0,
\end{equation}
which is solved by
\begin{equation}
    \delta \Omega_{112}=C_1(t)\text{cos} \left(\frac{2\sqrt{\Lambda}}{3}a(t)z\right) + C_2(t)\text{sin} \left(\frac{2\sqrt{\Lambda}}{3}a(t)z\right),
\end{equation}
where $C_1$ and $C_2$ are arbitrary functions of time. Concerning the vector components of the connection, instead, they turn to be identically vanishing as expected, i.e. $\delta P^\mu=\delta Q^\mu=\delta S^\mu=0$. At this point, it can be checked that the scalar field equation is automatically satisfied and the set of solutions is consistent. The remaining equations stemming from the linearization of \eqref{equation connection general} form a system of coupled equations relating the metric and the rank-3 torsion perturbations, given by
\begin{align}
   \frac{6}{\sqrt{\Lambda}} \delta q_{201}' +   \left( 2a \delta q_{202} - \sqrt{3}\delta q_{213} \right) &=  \left( \frac{3}{\sqrt{\Lambda}} \dot{h}_\times '-2\sqrt{3}h_\times'  - \frac{3}{2}a \dot{h}_+ + \sqrt{3\Lambda} a h_+ \right), \\
   \frac{6}{\sqrt{\Lambda}}  \delta q_{113}' +   \left( \sqrt{3}a^2\delta q_{202} + 2a \delta q_{213} \right) &=  \left( -\frac{3}{\sqrt{\Lambda}} h_+'' - \frac{\sqrt{3}}{2} a^2\dot{h}_+  + \sqrt{\Lambda}a^2 h_+ \right),\\
  \frac{6}{\sqrt{\Lambda}}   \delta q_{202}' +   \left( \sqrt{3} \delta q_{113} - 2a\delta q_{201} \right) &=   \left( -\frac{3}{\sqrt{\Lambda}} \dot{h}_+ ' + 2\sqrt{3} h_+'- \frac{3}{2} a\dot{h}_\times   + \sqrt{3\Lambda}a h_\times \right),\\
   \frac{6}{\sqrt{\Lambda}}  \delta q_{213}' -   \left(2a \delta q_{113} + \sqrt{3}a^2\delta q_{201} \right) &=   \left( -\frac{3}{\sqrt{\Lambda}} h_\times'' - \frac{\sqrt{3}}{2}a^2\dot{h}_\times  + \sqrt{\Lambda}a^2 h_\times \right).
\end{align}
The metric equations instead reduce to
\begin{align}
    \ddot{h}_+ -\frac{1}{a^2}h_+''-4\sqrt{\frac{\Lambda}{3}} \dot{h}_+  + \frac{4\Lambda}{3} h_+ &= 2 \left( \frac{1}{a^2}\delta q_{113}' - \dot{\delta q}_{202} \right) + \sqrt{\Lambda}\left( \sqrt{3} \delta q_{202} +\frac{1}{a} \delta q_{213} \right),\\
    \ddot{h}_\times -\frac{1}{a^2}h_\times''-4\sqrt{\frac{\Lambda}{3}} \dot{h}_\times  + \frac{4\Lambda}{3} h_\times &=2 \left( \frac{1}{a^2}\delta q_{213}' + \dot{\delta q}_{201} \right) - \sqrt{\Lambda}\left( \sqrt{3} \delta q_{201} +\frac{1}{a} \delta q_{113} \right),
\end{align}
where we used the fact that $\dot{a} = a \sqrt{\Lambda/3}$ and $\ddot{a} = a \Lambda/3$. In the adiabatic limit, where the scale factor can be considered nearly constant during the propagation of the wave signal, it is possible to analyze the above equations in Fourier space, where they boil down to the algebraic problem
\begin{align}
    & i k \delta q_{201} + \frac{\sqrt{\Lambda}}{6}\leri{2\delta  q_{202}-\sqrt{3}\delta q_{213}}=\frac{\sqrt{\Lambda}}{12}\left[ \left(3i\omega+2\sqrt{3\Lambda}\right)h_+ +k\left(\frac{6}{\sqrt{\Lambda}}\omega - 4i\sqrt{3}\right)h_\times\right], \\
    & i k \delta q_{113} + \frac{\sqrt{\Lambda}}{6}\leri{\sqrt{3}\delta q_{202}+2\delta q_{213}}=\frac{1}{2}\leri{k^2+i\frac{\omega}{2}\sqrt{\frac{\Lambda}{3}}+\frac{\Lambda}{3}}h_+, \\
    & i k \delta q_{202} + \frac{\sqrt{\Lambda}}{6}\leri{\sqrt{3}\delta q_{113}-2\delta  q_{201}}=\frac{\sqrt{\Lambda}}{12}\left[ \left(3i\omega+2\sqrt{3\Lambda}\right)h_\times -k\left(\frac{6}{\sqrt{\Lambda}}\omega - 4i\sqrt{3}\right)h_+ \right],\\
    & i k \delta q_{213} + \frac{\sqrt{\Lambda}}{6}\leri{-2\delta q_{113}-\sqrt{3}\delta q_{201}}=\frac{1}{2}\leri{k^2+i\frac{\omega}{2}\sqrt{\frac{\Lambda}{3}}+\frac{\Lambda}{3}}h_\times,
\end{align}
and
\begin{align}
    & \leri{\omega^2-k^2-\frac{4\Lambda}{3}-4i\sqrt{\frac{\Lambda}{3}}\omega}h_+=-2ik\,\delta q_{113}-(2i\omega+\sqrt{3\Lambda})\delta q_{202}-\sqrt{\Lambda}\,\delta q_{213},\\
    & \leri{\omega^2-k^2-\frac{4\Lambda}{3}-4i\sqrt{\frac{\Lambda}{3}}\omega}h_\times=-2ik\, \delta q_{213}+(2i\omega+\sqrt{3\Lambda})\delta q_{201}+\sqrt{\Lambda}\,\delta q_{113}.
\end{align}
In particular, we note that we can solve for the torsion perturbations in terms of the metric stresses, leading to:
\begin{align}
    &\delta q_{201}=\frac{p(k,\omega,\Lambda)h_+ -q(k,\omega,\Lambda)h_\times}{\Delta(k,\Lambda)},\\
    &\delta q_{113}=\frac{-m(k,\omega,\Lambda)h_+ +n(k,\omega,\Lambda)h_\times}{\Delta(k,\Lambda)},\\
    &\delta q_{202}=\frac{q(k,\omega,\Lambda)h_+ +p(k,\omega,\Lambda)h_\times}{\Delta(k,\Lambda)},\\
    &\delta q_{213}=-\frac{n(k,\omega,\Lambda)h_+ +m(k,\omega,\Lambda)h_\times}{\Delta(k,\Lambda)},
\end{align}
where we introduced
\begin{align}
    \Delta(k,\Lambda)&\equiv 2(1296 k^4-72 \Lambda k^2+49\Lambda^2),\\
    p(k,\omega,\Lambda)&\equiv 2 k \sqrt{\Lambda } \left(108 k^2 \omega +2 i
   \sqrt{3} \Lambda ^{3/2}-3 \Lambda  \omega
   \right),\\
    q(k,\omega,\Lambda)&\equiv3 \left(72 k^4 \left(5 \sqrt{3} \sqrt{\Lambda
   }+6 i \omega \right)-6 k^2 \left(3 \sqrt{3}
   \Lambda ^{3/2}+8 i \Lambda  \omega \right)+7
   \Lambda ^2 \left(2 \sqrt{3} \sqrt{\Lambda }+3
   i \omega \right)\right),\\
    m(k,\omega,\Lambda)&\equiv 12 i k \left(108 k^4-3 k^2 \Lambda +2 i \sqrt{3}
   \Lambda ^{3/2} \omega +4 \Lambda ^2\right),\\
    n(k,\omega,\Lambda)&\equiv\sqrt{\Lambda } \left(432 k^4-12 k^2
   \left(\Lambda -3 i \sqrt{3} \sqrt{\Lambda }
   \omega \right)+7 i \sqrt{3} \Lambda ^{3/2}
   \omega +14 \Lambda ^2\right).
\end{align}
The solutions for the torsion can be then rewritten in the compact matrix form
\begin{equation}
    \mathbf{Q} = M
    \mathbf{H},
    \label{matrix tensor torsion}
\end{equation}
where the matrix $M$ and the vectors $\mathbf{Q}, \mathbf{H}$ are defined as
\begin{equation}
\mathbf{Q}=
\begin{pmatrix}
\delta q_{201}\\
\delta q_{113}\\
\delta q_{202}\\
\delta q_{213}
\end{pmatrix},\;
M=\frac{1}{\Delta}
\begin{pmatrix}
p & -q\\
-m & n \\
q & p\\
-n & -m
\end{pmatrix},\;
\mathbf{H}=
\begin{pmatrix}
h_{+}\\
h_{\times}\\
\end{pmatrix}.
\end{equation}
We observe that also the equations for the metric perturbations admit an analogous formulation, given by
\begin{equation}
    d(k,\omega,\Lambda) \mathbf{H}=N \mathbf{Q},
\end{equation}
where in this case the matrix $N$ has the form 
\begin{equation}
    N=
    \begin{pmatrix}
    0 & \quad-2 i k &\quad - (2 i \omega+\sqrt{3\Lambda}) &\quad - \sqrt{\Lambda}\\
    2 i \omega+\sqrt{3\Lambda} & \quad\sqrt{\Lambda} &\quad 0 &\quad - 2i k
    \end{pmatrix}
\end{equation}
and the function $d(k,\omega,\Lambda)$ is defined as
\begin{equation}
    d(k,\omega,\Lambda)\equiv\omega^2-k^2-\frac{4\Lambda}{3}-4i\sqrt{\frac{\Lambda}{3}}\omega.
\end{equation}
Then, by taking into account \eqref{matrix tensor torsion}, the equation for the metric can be rewritten as
\begin{equation}
    \leri{d(k,\omega,\Lambda)\,I-P} \mathbf{H}=0,
    \label{matrix metric}
\end{equation}
where $I$ is the identity matrix of dimension two and we used the associativity property of the matrix product for defining the square matrix $P=N M$, given by:
\begin{equation}
    P=
    \begin{pmatrix}
    p_{11} & -p_{12}\\
    p_{12} & p_{11}
    \end{pmatrix},
\end{equation}
with elements
\begin{equation}
    p_{11}\equiv\frac{ 2ikm-(2i\omega+\sqrt{3\Lambda})q+\sqrt{\Lambda}n}{\Delta},\qquad p_{12}\equiv\frac{ 2ikn+(2i\omega+\sqrt{3\Lambda})p-\sqrt{\Lambda}m}{\Delta}.
\end{equation}
It follows that the equations for the metric perturbations take the form (see \eqref{matrix metric}):
\begin{align}
    (d(k,\omega,\Lambda)-p_{11})h_+ + p_{12}h_\times&=0,\\
    (d(k,\omega,\Lambda)-p_{11})h_\times - p_{12}h_+&=0.
\end{align}
We immediately see that when $p_{12}=0$ the tensor modes do not mix each other and the cross and plus polarizations simply retain the same dispersion relation given by
\begin{equation}
    \mathcal{D}(k,\omega)\equiv d(k,\omega,\Lambda)-p_{11}=0.
    \label{dispersion no birifrengence}
\end{equation}
In the general case, however, we deal with $p_{12}\neq 0$ and it is then useful to introduce the left and right handed polarization states, i.e.
\begin{equation}
    h_L=\frac{1}{\sqrt{2}}(h_+ -i h_\times),\quad h_R=\frac{1}{\sqrt{2}}(h_+ +i h_\times),
\end{equation}
which allows us to decouple \eqref{matrix metric}, leading to
\begin{align}
    &(d(k,\omega,\Lambda)-p_{11}+ip_{12})h_L=0,\\
    &(d(k,\omega,\Lambda)-p_{11}-ip_{12})h_R=0,
\end{align}
from which it is possible to obtain the dispersion relations
\begin{equation}
    \mathcal{D}_{L,R}(k,\omega)\equiv d(k,\omega,\Lambda)-p_{11}\pm i p_{12}=0.
\label{dispersion relation}
\end{equation}
They imply the emergence of birefringence as in the metric formulation of CSMG (see \cite{Martin-Ruiz:2017cjt,Nojiri:2019nar,Nojiri:2020pqr} for a comparison), with the asymmetry quantified by the parameter $p_{12}$. Solving for $\omega$ yields:
\begin{align}
    \omega_L &= \pm\frac{\sqrt{36 k^4-12 k^3 \sqrt{\Lambda }-35 k^2 \Lambda -18 k \Lambda ^{3/2}-3 \Lambda ^2} }{2 \left(3 k+\sqrt{\Lambda }\right)}+i\sqrt{\frac{\Lambda}{12}},
    \label{omega left}
    \\
   \omega_R &= \pm\frac{ \sqrt{36 k^4+12 k^3 \sqrt{\Lambda }-35 k^2 \Lambda +18 k \Lambda ^{3/2}-3 \Lambda ^2} }{2 \left(3 k-\sqrt{\Lambda }\right)}+i\sqrt{\frac{\Lambda}{12}},
   \label{omega right}
\end{align}
and we immediately identify in \eqref{omega left}-\eqref{omega right} the cosmological damping due to the cosmological constant, without any difference between the left and the right polarization. In addition, we observe that in order to have the propagation of wave perturbations we have to require
\begin{equation}
    36 k^4\pm12 \sqrt{\Lambda } k^3-35 \Lambda k^2  \pm18 \Lambda ^{3/2}k -3 \Lambda ^2>0,
\end{equation}
leading to
\begin{align}
    &k<0,\;0<\Lambda<9k^2;\qquad k>0,\;0<\Lambda< \gamma k^2,
    \label{condition propagation1}
\end{align}
for the left mode and
\begin{align}
    &k>0,\;0<\Lambda<9k^2;\qquad k<0,\;0<\Lambda< \gamma k^2,
    \label{condition propagation2}
\end{align}
for the right mode, where the parameter $\gamma$ is 
\begin{equation}
    \gamma=\frac{1}{9} \left(\sqrt[3]{2592 \sqrt{62}-6697}-\frac{719}{\sqrt[3]{2592 \sqrt{62}-6697}}+11\right)\sim 0.543988.
\end{equation}
Then, since in the adiabatic limit $|k|\gg\sqrt{\frac{\Lambda}{3}}$, from \eqref{condition propagation1}-\eqref{condition propagation2} one always has wave propagation and birefringence occurs for all the wave vectors under consideration. Given \eqref{omega left}-\eqref{omega right}, we display the expressions of the group and phase velocity for the circularly polarized modes:
\begin{align}
    v^g_{L,R}&\equiv\frac{d \omega_{L,R}}{d k}=\dfrac{108k^4\pm 54 k^3 \sqrt{\Lambda}-18k^2 \Lambda\mp 8 k \Lambda^\frac{3}{2}}{2\leri{ 3k \pm \sqrt{\Lambda}}^2\sqrt{36k^4 \mp12 k^3 \sqrt{\Lambda}-35 k^2 \Lambda \mp 18 k \Lambda^\frac{3}{2}-3\Lambda^2}}, \\
    v^p_{L,R}&\equiv\frac{ \omega_{L,R}}{k}=\dfrac{\sqrt{36 k^4 \mp 12 k^3 \sqrt{\Lambda }-35 k^2 \Lambda \mp 
    18 k \Lambda ^{3/2}-3 \Lambda ^2}}{2k\leri{3k\pm\sqrt{\Lambda}}}.
\end{align}
It is useful to expand these quantities in a power series of the parameter $\epsilon \equiv \frac{\sqrt{\Lambda}}{k}$ which, in the adiabatic limit, can be taken as $|\epsilon|\ll 1$. We obtain
\begin{align}
    v^g_{L,R}&=1+\dfrac{\epsilon^2}{3}\pm \dfrac{4\epsilon^3}{9}+\mathcal{O}\leri{\epsilon^4},\\
    v^p_{L,R}&=1\mp\dfrac{\epsilon}{2}-\dfrac{\epsilon^2}{3}\mp \dfrac{2\epsilon^3}{9}+\mathcal{O}\leri{\epsilon^4}.
\end{align}
These results are in agreement with the discussion in \cite{Wu:2021ndf} and, as expected, the group velocity turns out to be greater than the vacuum speed of light and independent upon the type of polarization at order $\mathcal{O}(\epsilon^2)$. From these expressions it is clear, therefore, that a measurement of the phase velocity of gravitational waves is much more efficient in resolving the two polarizations states, given that $|v^p_L-v^p_R|\sim \mathcal{O}\leri{\epsilon}$, whereas in the case of a group velocity detection it is expected $|v^g_L-v^g_R|\sim \mathcal{O}\leri{\epsilon^3}$. The deviation of the group velocity of both polarizations from the speed of light can be, in principle, compared with the current constraint on gravitational waves speed derived in
\cite{LIGOScientific:2017zic}, which reads
\begin{equation}\label{boundvg}
    -3 \cross 10^{-15} \leq v_g -1 \leq 7 \cross 10^{-16}.
\end{equation}
If we assume a magnitude of the wavenumber $k=10^{-7} \, \text{m}^{-1}$, corresponding to a frequency $\nu \approx 50 \, \text{Hz}$ well inside the sensitivity curves of ground-based interferometers, and the measured value of the cosmological constant $\Lambda_{\text{exp}} \approx 10^{-52}\, \text{m}^{-2}$ we calculate a theoretically expected deviation  $\mathcal{O}\leri{\epsilon^2}=10^{-38}$. Therefore the bound \eqref{boundvg} is not sufficiently tight in order to falsify this model of modified gravity. However, the dependence of the deviation parameter with respect to the wavenumber, i.e. $\epsilon^2 \propto k^{-2}$, indicates that for detections in the low-frequencies domain the expected deviation would result significantly larger. By considering a signal in the mHz band, which will become accessible with the space interferometer LISA, we obtain a much greater expected deviation  $\mathcal{O}\leri{\epsilon^2}=10^{-29}$. The maximum magnitude for the deviation parameter is reached in the case of a nHz gravitational wave, detectable with pulsar timing arrays, which we calculate to be $\mathcal{O}\leri{\epsilon^2}=10^{-17}$.

\subsection{Scalar modes}
To study the evolution of the scalar sector we consider the scalar perturbation $\delta\theta$ together with the scalar modes of the metric \cite{Nojiri:2020pqr}, which for the gauge choice of sec.~\ref{sec6} are encoded in the trace $h=h\indices{^i_i}$ and a traceless contribution depending on gradients of a function $B$, i.e.
\begin{equation}
    h_{ij} = \frac{1}{3}\bar{g}_{ij} h + \partial_i \partial_j B -\frac{1}{3}\bar{g}_{ij} \partial^k\partial_k B.
\end{equation}
We expect that these scalar modes couple to components of affine perturbations that behave as scalars under spatial rotations, that is to say the component $\delta\Omega_{000}$ of the tensor part of nonmetricity and the time component of the torsion and nonmetricity vectors, i.e. $\delta S^t$, $\delta P^t$ and $\delta Q^t$. We note that also contributions of the form $\delta \Omega_{0ij}=\psi \delta_{ij}$ would in principle be allowed, but they are eventually ruled out by the traceless character of $\delta\Omega_{\mu\nu\rho}$ and $\delta q_{\mu\nu\rho}$. Therefore, assuming again a wave propagating along the z-axis, we find the following solution for the metric perturbation:
\begin{align}
    h_{xx} &= h_{yy} =  a^2 \left( -\frac{c_1}{3\sqrt{3\Lambda}} e^{-\sqrt{3\Lambda}t} + \frac{c_2}{3} \right),\\
    h_{zz} &=a^2 \left( -\frac{c_1}{3\sqrt{3\Lambda}} e^{-\sqrt{3\Lambda}t} + \frac{c_2}{3} + f(z) \right),
\end{align}
so that
\begin{equation}
    h = -\frac{c_1}{\sqrt{3\Lambda}} e^{-\sqrt{3\Lambda}t} + c_2 + f(z).
\end{equation}
The pseudo-scalar perturbation is instead given by
\begin{equation}
    \delta\theta = -\frac{c_1}{6\Lambda^{3/2}} e^{-\sqrt{3\Lambda}t} + \frac{c_2}{2\sqrt{3}\Lambda} + c_0,
\end{equation}
while for the affine perturbations we have $\delta\Omega_{000}=0$ and
\begin{align}
    \delta S^t &= \frac{c_1}{2\sqrt{3}}e^{-\sqrt{3\Lambda}t},\\
    \delta Q^t &= 4\delta P^t=  -\frac{5c_1}{3}e^{-\sqrt{3\Lambda}t}.
\end{align}
In the above expressions $c_0$, $c_1$, $c_2$ are constants of integration and $f(z)$ is an arbitrary function of $z$. We observe, therefore, that in agreement with the predictions of the purely metric formulation of CSMG, also in this case scalar polarizations do not propagate as a wave, but they decay exponentially in time over a time scale $t_D$ of order $t_D\sim 1/3H_0$, so that in the adiabatic limit they can be considered nearly constant with respect to the tensor polarizations.

\section{Summary and discussion}\label{sec7}
In this work we considered projective invariant metric-affine Chern-Simons gravity, where the general relativity action is modified by including a parity violating term coupled to a pseudo-scalar field $\theta(x)$. Projective symmetry is recovered via the inclusion of an additional nonmetricity contribution to the usual Pontryagin density, assuring the absence of instabilities of the same nature as in \cite{BeltranJimenez2019}.

Dynamical instabilities can also arise in the presence of third order derivatives of the metric in the field equations. While in metric CSMG they enter directly the metric equations via the C-tensor, in the metric-affine formulation they can only arise on half-shell (once the solution for the connection is considered), from terms featuring first derivatives of the affine components. In particular, in the metric-affine theories one has the option of restricting the analysis to a specific subclass of the most general metric-affine geometry (Weyl, torsionless or metric compatible geometries, for instance) in such a way that possibly dangerous terms are a priori absent. Regarding metric-affine CSMG, when the matter sector does not couple to the connection, we proved that these terms can only arise from the 3-rank tensor components of nonmetricity and torsion, so that when these are vanishing, the absence of higher derivatives of the metric in the field equations is guaranteed.

If instead the tensor components are present, as we assumed in deriving our results, then one has to resort to the same solutions considered in metric CSMG, i.e. consider the theory in the small coupling limit or exploiting specific symmetries of the solutions to get rid of terms introducing higher-order derivatives.\\
A natural setting where this happens is provided by the isotropic and homogeneous FLRW spacetime. In such a cosmological setting, one can exploit the spacetime symmetries in order to constrain the most general form of the affine connection. On one hand, this immediately implies that the tensor part of torsion is vanishing. On the other hand, it allows to greatly simplify the connection field equations reducing them to algebraic equations. This allows to obtain exact solutions for torsion and nonmetricity, which are only characterized by their vector components, ultimately sourced by the scale factor and time derivatives of the scalar field. In particular, we obtain the condition of vanishing tensor part of nonmetricity as a solution to the equations. Summing up, although we consider the most general affine sector, including a priori the tensor components of torsion and nonmetricity, we obtain the vanishing of the former imposing the FRW symmetries and of the latter as a dynamical condition solving the equations, thus guaranteeing the absence of higher order derivatives in the metric field equations.

Reinserting the solutions back into the remaining equations, one eventually obtains three independent conditions that can be solved yielding different profiles for the scale factor and the scalar field. We reported three different solutions. Two of them feature power law and linear behaviors for $a(t)$, with the scalar field growing quadratically in time. In particular, the power law behavior allows to reproduce the well-known radiation and matter dominated scenarios and it is characterized by a negative bare energy density $\rho_0$, which acquires a correction due to the additional energy contribution of torsion and nonmetricity, resulting in a positive total energy density and a standard background expansion.
The same holds for the third solution, describing a de Sitter phase of expansion, where $\rho_0 = -\Lambda /2\kappa^2$ but $\tilde{\rho} = \Lambda/\kappa^2$. In this solution, the scalar field grows linearly, playing the role of a cosmological time, in agreement with \cite{Nojiri:2019nar}.

We then proceeded to study gravitational wave propagation on the de Sitter background. We first focused on purely tensor modes, neglecting the scalar field perturbation, considering a transverse and traceless metric perturbation and perturbing all affine components as well. Consistently, the perturbations of the tensorial components, $\delta q_{\mu\nu\rho}$ and $\delta \Omega_{\mu\nu\rho}$ are the only nonvanishing ones. However, nonmetricity perturbations are decoupled from metric ones and behave as a harmonic oscillator, without propagating. Torsion perturbations instead affect the propagation of the gravitational wave, modifying its dispersion relation and introducing parity violating effects. These are apparent once the wave is decomposed into left and right handed circular polarizations. The phase and group velocities can be computed in the adiabatic approximation and they are both deviating from the speed of light in vacuum. The sign of the correction depends on the chirality of the wave, so that left and right handed modes travel at different speeds. Thus, we establish the existence of velocity birefringence in the propagation of gravitational waves in metric-affine Chern-Simons theory.

The deviation in the group velocity can be compared to the current bound on the speed of gravitational waves. However, for a wave in the frequency band of current detectors and using the measured value for the cosmological constant we conclude that the experimental bound is not sufficiently tight to constrain the theory.

Also, current experimental constraints on gravitational wave birefringence \cite{Zhao_2022} have only been used to constrain models producing a $\sim k^3$ correction into the dispersion relation, i.e. $\sim k$ correction to the phase velocity. However, in the present case the phase velocity is modified by $\sim k^{-1}$ terms, and the bounds estimated in \cite{Zhao_2022} can not be straightforwardly applied, but it deserves further analysis. The friction term, instead, is the same for both polarizations in the metric wave equation, and it is provided by the damping effect due to the universe expansion, so that parity violating effects leading to amplitude birefringence in vacuum are absent. However, as discussed in \cite{Bombacigno:2022naf}, such a phenomenon can arise when propagation in matter is addressed, by virtue of the kinematic damping experienced by the wave when interacting with the particles of the traversed medium.

Finally, we analysed the scalar sector of the perturbations, consisting in the pseudo-scalar field $\delta\theta$ and the scalar modes contained in the metric tensor and affine components.  Scalar perturbations do not carry any parity violating signature nor observable effects since the scalar modes are not propagating but they rather exponentially decay in time.

\acknowledgments
The work of F.B is supported by the postdoctoral grant CIAPOS/2021/169. The work of F. M. is supported by the Della Riccia foundation grant for the year 2022. This work is supported by the Spanish Grant FIS2017-84440-C2- 1-P funded by MCIN/AEI/10.13039/5011\\
00011033 “ERDF A way of making Europe”, Grant PID2020-116567GB-C21 funded by MCIN/AEI/10.13039/501100011033, the project PROMETEO/2020/079 (Generalitat Valenciana), and by the European Union's Horizon 2020 research and innovation programme under the H2020-MSCA-RISE-2017 Grant No. FunFiCO-777740.

\appendix
\section{Metric-affine formalism}\label{appendix A}
\noindent In this appendix we review some basic notions about the metric-affine formalism we adopted throughout the paper. The Riemann tensor is defined in terms of the independent connection as:
\begin{equation}
    \mathcal{R}\indices{^\rho_{\mu\sigma\nu}}=\partial_\sigma\Gamma\indices{^\rho_{\mu\nu}}-\partial_\nu\Gamma\indices{^\rho_{\mu\sigma}}+\Gamma\indices{^\rho_{\tau\sigma}}\Gamma\indices{^\tau_{\mu\nu}}-\Gamma\indices{^\rho_{\tau\nu}}\Gamma\indices{^\tau_{\mu\sigma}},
\end{equation}
and covariant derivatives act as
\begin{equation}
    \nabla_\mu T\indices{^\rho_\sigma}=\partial_\mu T\indices{^\rho_\sigma}+\Gamma\indices{^\rho_{\lambda\mu}}T\indices{^\lambda_\sigma}-\Gamma\indices{^\lambda_{\sigma\mu}}T\indices{^\rho_\lambda}.
\end{equation}
We are considering the affine connection as general as possible, so that we can introduce torsion and nonmetricity tensors, which read respectively:
\begin{equation}
    \begin{split}
        &T\indices{^\rho_{\mu\nu}}\equiv\Gamma\indices{^\rho_{\mu\nu}}-\Gamma\indices{^\rho_{\nu\mu}},\\
        &Q\indices{_{\rho\mu\nu}}\equiv-\nabla_\rho g_{\mu\nu}.
    \end{split}
\end{equation}
In evaluating the equation of motion for the connection from \eqref{equation connection general}, we used the generalized Palatini identity
\begin{equation}
    \delta\mathcal{R}\indices{^\rho_{\mu\sigma\nu}}=\nabla_\sigma\delta\Gamma\indices{^\rho_{\mu\nu}}-\nabla_\nu\delta\Gamma\indices{^\rho_{\mu\sigma}}-T\indices{^\lambda_{\sigma\nu}}\delta\Gamma\indices{^\rho_{\mu\lambda}},
\end{equation}
and the property for vector densities
\begin{equation}
    \int d^4 x\; \nabla_\mu\leri{\sqrt{-g} V^\mu}=\int d^4x\; \partial_\mu\leri{\sqrt{-g}V^\mu}+\int d^4x\; \sqrt{-g}\;T\indices{^\rho_{\mu\rho}} V^\mu=\int d^4x\; \sqrt{-g}\;T\indices{^\rho_{\mu\rho}} V^\mu.
\end{equation}
We also rewrote torsion and nonmetricity in their irreducible parts:
\begin{align}
    &T_{\mu\nu\rho} = \dfrac{1}{3}\left(T_{\nu}g_{\mu\rho}-T_{\rho}g_{\mu\nu}\right) +\dfrac{1}{6} \varepsilon_{\mu\nu\rho\sigma}S^{\sigma} + q_{\mu\nu\rho},\label{torsion decomposition}\\
    &Q_{\rho\mu\nu}=\frac{5Q_\rho-2P_\rho}{18}g_{\mu\nu}-\frac{Q_{(\mu}g_{\nu)\rho}-4P_{(\mu}g_{\nu)\rho}}{9}+\Omega_{\rho\mu\nu}.
    \label{non metricity decomposition}
\end{align}
In particular, we introduced the trace vector
\begin{equation}
T_{\mu} \equiv T \indices{^{\nu}_{\mu\nu}},
\end{equation}
the pseudotrace axial vector
\begin{equation}
S_{\mu} \equiv \varepsilon_{\mu\nu\rho\sigma}T^{\nu\rho\sigma},
\end{equation}
and the antisymmetric tensor $q_{\mu\nu\rho}=-q_{\mu\rho\nu}$ satisfying
\begin{equation}
\varepsilon^{\mu\nu\rho\sigma} q_{\nu\rho\sigma} = 0, \qquad q\indices{^{\mu}_{\nu\mu}} = 0.
\end{equation}
While for what concerns the nonmetricity, we defined the Weyl vector
\begin{equation}
    Q_\rho=Q\indices{_\rho^\mu_\mu},
\end{equation}
the second trace
\begin{equation}
    P_\rho=Q\indices{^\mu_{\mu\rho}}=Q\indices{^\mu_{\rho\mu}},
\end{equation}
and the traceless part $\Omega_{\rho\mu\nu}$, obeying
\begin{equation}
    \Omega_{\rho\mu\nu}=\Omega_{\rho\nu\mu}.
\end{equation}
It is always possible, moreover, to rewrite the affine connection as
\begin{equation}
    \Gamma\indices{^\rho_{\mu\nu}}=L\indices{^\rho_{\mu\nu}}+N\indices{^\rho_{\mu\nu}}=L\indices{^\rho_{\mu\nu}}+K\indices{^\rho_{\mu\nu}}+D\indices{^\rho_{\mu\nu}},
    \label{christoffel contorsion disformal}
\end{equation}
where $L\indices{^\rho_{\mu\nu}}$ denotes the Christoffel symbols and the contorsion and disformal tensors are given by, respectively
\begin{align}
    &K\indices{^\rho_{\mu\nu}}=\frac{1}{2}\leri{T\indices{^\rho_{\mu\nu}}-T\indices{_\mu^\rho_\nu}-T\indices{_\nu^\rho_\mu}}=-K\indices{_\mu^\rho_{\nu}},
    \label{decomposition contorsion}\\
    &D\indices{^\rho_{\mu\nu}}=\frac{1}{2}\leri{Q\indices{_{\mu\nu}^\rho}+Q\indices{_{\nu\mu}^\rho}-Q\indices{^\rho_{\mu\nu}}}=D\indices{^\rho_{\nu\mu}}.
    \label{decomposition disformal}
\end{align}
For a generic metric-affine structure the Riemann tensor is skew-symmetric only in its last two indices, so that we can in principle take the different traces
\begin{align}
    \mathcal{R}_{\mu\nu}&\equiv\mathcal{R}\indices{^\alpha_{\mu\alpha\nu}},\\
    \hat{\mathcal{R}}_{\mu\nu}&\equiv \mathcal{R}\indices{^\alpha_{\alpha\mu\nu}}=\partial_{[\mu}Q_{\nu]},\\
    \mathcal{R}^\dag_{\mu\nu}&\equiv g_{\mu\tau}g^{\rho\sigma}\mathcal{R}\indices{^\tau_{\rho\sigma\nu}},
\end{align}
which are called respectively Ricci, homothetic curvature and co-Ricci tensors. In terms of the distorsion tensor the Riemann curvature can be rewritten as
\begin{equation}
 \mathcal{R}_{\mu\rho\nu\sigma}=R_{\mu\rho\nu\sigma}+\tensor[^{(L)}]{\nabla}{_{\nu}}N_{\mu\rho\sigma}-\tensor[^{(L)}]{\nabla}{_{\sigma}} N_{\mu\rho\nu}+N_{\mu\lambda\nu}N\indices{^\lambda_{\rho\sigma}}-N_{\mu\lambda\sigma}N\indices{^\lambda_{\rho\nu}},
    \label{perturbative expansion riemann}
\end{equation}
where $R\indices{^\mu_{\nu\rho\sigma}}$ and $\tensor[^{(L)}]{\nabla}{_{\mu}}$ are built from the Levi Civita connection. The affine Einstein tensor $\mathcal{G}_{\mu\nu}$ can be therefore rewritten as
\begin{equation}
    \mathcal{G}_{\mu\nu}=G_{\mu\nu}+C_{\mu\nu},
\end{equation}
where we introduced the C-tensor \begin{equation}
\begin{split}
    C_{\mu\nu}\equiv& \tensor[^{(L)}]{\nabla}{_{\rho}}N\indices{^\rho_{(\mu\nu)}}-\tensor[^{(L)}]{\nabla}{_{(\nu}}N\indices{^\rho_{\mu)\rho}}+N\indices{^\rho_{\lambda\rho}}N\indices{^\lambda_{(\mu\nu)}}-N\indices{^\rho_{\lambda(\nu}}N\indices{^\lambda_{\mu)\rho}}+\\
    &-\frac{1}{2}g_{\mu\nu}\leri{\tensor[^{(L)}]{\nabla}{_{\rho}}N\indices{^{\rho\sigma}_\sigma}-\tensor[^{(L)}]{\nabla}{_{\sigma}}N\indices{^{\rho\sigma}_{\rho}}+N\indices{^\rho_{\lambda\rho}}N\indices{^{\lambda\sigma}_\sigma}-N\indices{^\rho_{\lambda\sigma}}N\indices{^{\lambda\sigma}_\rho}}.
\end{split}
\label{c tensor affine}
\end{equation}
It is worth mentioning that as opposed to the metric formulation, in this case the C-tensor is not traceless, i.e.
\begin{equation}
    C=C\indices{^\mu_\mu}=-\tensor[^{(L)}]{\nabla}{_{\rho}}N\indices{^{\rho\sigma}_\sigma}+\tensor[^{(L)}]{\nabla}{_{\sigma}}N\indices{^{\rho\sigma}_{\rho}}-N\indices{^\rho_{\lambda\rho}}N\indices{^{\lambda\sigma}_\sigma}+N\indices{^\rho_{\lambda\sigma}}N\indices{^{\lambda\sigma}_\rho}.
\label{c trace}
\end{equation}
This implies that vacuum solutions of CSMG are not necessarily characterized by a vanishing Ricci scalar as in GR, since now the following holds
\begin{equation}
    R=C-\kappa T,
\end{equation}
which allows to recast the equations for the metric in the form
\begin{equation}
 R_{\mu\nu}=\kappa T_{\mu\nu}-C_{\mu\nu}-\frac{1}{2}g_{\mu\nu}\leri{\kappa T-C}.
\end{equation}

\bibliographystyle{JHEP}
\bibliography{references}
\end{document}